# Cavity optomechanical magnetometer actuated by substrate bending

D. D. Florizone,[1] I. E. Rabuka[1], K. D. Hiltz[1], I. Grobbellaar[1], E. F. DeCorby[2], A. Chen[2], K. G. Scheuer,[2] and R. G. DeCorby[1,2,a)]

[1] *ECE Department, University of Alberta, 9211-116 St. NW, Edmonton, AB, Canada, T6G 1H9*

[2] *Ultracoustics Technologies Ltd., 10230 Jasper Ave. NW, Edmonton, AB T5J 4P6, Canada*

We describe a magnetic field sensor based on a monolithic Fabry-Perot cavity. The cavities were fabricated in a thin-film buckling process at one surface of a fused silica chip, and the chip was subsequently bonded to a giant magnetostrictive material (GMM). External magnetic fields drive flexural modes of the assembly and the associated bending stresses are efficiently mapped to changes in the cavity height, with enhanced response at kHz-range mechanical resonance frequencies. A key attribute of this platform is that the size of the transducing material is decoupled from the size of the optical cavity. The buckled cavity operates in low mode order and with moderate $Q \sim 10^3$, providing a high optomechanical coupling coefficient while also enabling a straightforward tuned-to-slope readout technique. Peak sensitivity ~ 2 pT/Hz$^{1/2}$ was achieved in an unshielded laboratory environment, limited by laser phase and electronics noise in the readout system. With future reductions in these noise sources, we predict that a broadband thermomechanical-noise-limited sensitivity in the ~ 0.1 pT/Hz$^{1/2}$ range is feasible.

[a)] Author to whom correspondence should be addressed. Electronic mail: rdecorby@ualberta.ca

Magnetometers have important applications in medicine, surveying, navigation, and defense [1-4]. Performance metrics include the minimum detectable field, frequency range, bandwidth, dynamic range, spatial resolution, and directional dependence (i.e., scalar versus vector magnetometers). Practical considerations include the operational temperature range, and the overall size, weight and power (SWaP) associated with both the sensor and its control and readout instrumentation. Because of greatly differing cost and performance requirements, especially regarding sensitivity and frequency range, there is no 'one size fits all' magnetometer [4].

Ultra-sensitive devices, capable of detecting fields at or below picoTesla levels [1], are needed in the defense and security sectors (e.g., for detection of magnetic anomalies or unexploded ordnance), for geo-surveying, for non-destructive testing (NDT), and in bio-medicine. This space has traditionally been dominated by SQUID magnetometers [2], which can provide sensitivity as high as a few $fT/Hz^{1/2}$, and can be designed to operate within a wide frequency range (DC to GHz). However, their need for cryogenic cooling hinders mobility and limits their usage. Atomic vapor based magnetometers [5] have achieved sensitivity as high as ~ 160 $aT/Hz^{1/2}$ (i.e., for SERF magnetometers [2-5]) but require a heated gas vapor cell and often have limited frequency and dynamic ranges. Portable, commercial instruments can nevertheless provide impressive performance, with sensitivity ~ 1 $pT/Hz^{1/2}$ and bandwidth spanning the ~ DC to 500 Hz range for rubidium-based units [6]. More recently, solid-state magnetometers based on NV-defect centers in diamond have attracted interest [2,4]. Commercial instruments with sufficiently low SWaP to enable mobile deployment on UAVs reportedly achieve sensitivity ~ 100 $pT/Hz^{1/2}$ in the ~ DC to 500 Hz frequency range [7]. Challenges include the need for moderately high-power (e.g., Watts-level [8,9]) laser and microwave sources for state preparation and readout.

Cavity optomechanical magnetometers (COMs) [10-17] typically combine a giant magnetostrictive material (GMM) with a high-quality optomechanical resonator [18]. Low-power laser readout (i.e., at μW to mW levels) confers inherent SWaP and EMI-immunity advantages, and sub-$pT/Hz^{1/2}$ sensitivity has already been achieved [14,15]. Moreover, when operating at their thermomechanical noise limit [18-20], they can (in principle) provide wideband, frequency-independent sensitivity. These attributes, combined with their high dynamic range, make them intriguing candidates for ultra-sensitive detection in unshielded, non-cryogenic environments.

To date, peak sensitivity has been achieved only at one or more mechanical resonance frequencies [14]. The earliest works employed small GMM particles or thin films coupled to micron-scale whispering-gallery-mode (WGM) cavities, yielding low-order mechanical resonances in the MHz region. Notably, Hu et al. [16] reported a sensitivity of ~ 1.7 pT/Hz$^{1/2}$ at ~ 9.5 MHz, with broadband sensitivity < 1 nT/Hz$^{1/2}$ over the ~ 1 – 30 MHz range. Sensing in this range is of interest for NDT applications, as well as for studies of high-frequency circuits, biological entities, and fundamental physics of spin [12]. Nevertheless, most applications for magnetometers require high sensitivity in the Hz to kHz range. Motivated by this, more recent work has employed centimeter-scale GMMs to increase the magnetic sensing volume and to place low-order mechanical resonance frequencies in the kHz region [12-15]. Of note, a micro-optic Fabry-Perot assembly [14] achieved peak sensitivity ~ 0.62 pT/Hz$^{1/2}$ at ~ 5 kHz with sensitivity < 1 nT/Hz$^{1/2}$ in the ~ 2 – 70 kHz range, and a fiber-optic Fabry-Perot assembly [15] achieved a peak sensitivity of ~ 0.54 pT/Hz$^{1/2}$ at ~ 11 kHz.

Most of the previously reported COMs were characterized by sub-optimal coupling of the GMM strain to the mechanical modes, exacerbated by a relatively low optomechanical coupling coefficient (i.e., $G_\lambda = d\lambda/dx$, where $\lambda$ is wavelength and $x$ is mechanical displacement). To compensate, those systems employed ultra-high-$Q$ optical resonators (e.g., $Q > 10^7$ [12-14]), which necessitates the use of sophisticated stabilization and readout schemes such as the Pound-Drever-Hall technique [14].

We describe a COM based on a Fabry-Perot cavity fabricated using a thin-film buckling process (hereafter called a 'buckled dome cavity'). These cavities have been extensively discussed as ultrasound sensors in our previous work [19,21]. Briefly, they are fabricated by embedding a patterned, low-adhesion (Teflon-like) layer between a pair of quarter-wave-stack (QWS) mirrors, and inducing the compressively stressed upper mirror to delaminate and buckle over the regions of low adhesion, yielding half-symmetric Fabry-Perot resonators. The upper mirror also functions as a clamped-shell mechanical oscillator with its lowest vibrational resonances lying in the few MHz range. They operate in a low optical mode order (typically $1 < m < 3$) with moderate $Q \sim 10^3$ - $10^4$, provide a high optomechanical coupling coefficient ($G_\lambda \sim 1$), enable high-efficiency coupling to Gaussian optical beams, and achieve nearly thermal-noise-limited displacement sensitivity over a large bandwidth when read out using a straightforward tuned-to-slope technique and for laser interrogation powers < 100 μW. Additional details on the particular sensors employed here are provided in the Supplementary Information (SI) file.

We recently demonstrated that these cavities are sensitively coupled to the flexural modes of their underlying substrate [22]. Bending of the substrate, even driven by thermal energy at room temperature, modulates the buckled dome cavity height. Here, we exploit this effect to implement a broadband and ultrasensitive magnetometer by affixing the sensor substrate to a GMM plate as depicted in Fig. 1. The relatively large mass and volume of the magnetic material provides a large coupling to external magnetic fields, while the small size and high quality of the embedded Fabry Perot cavity results in a high readout sensitivity.

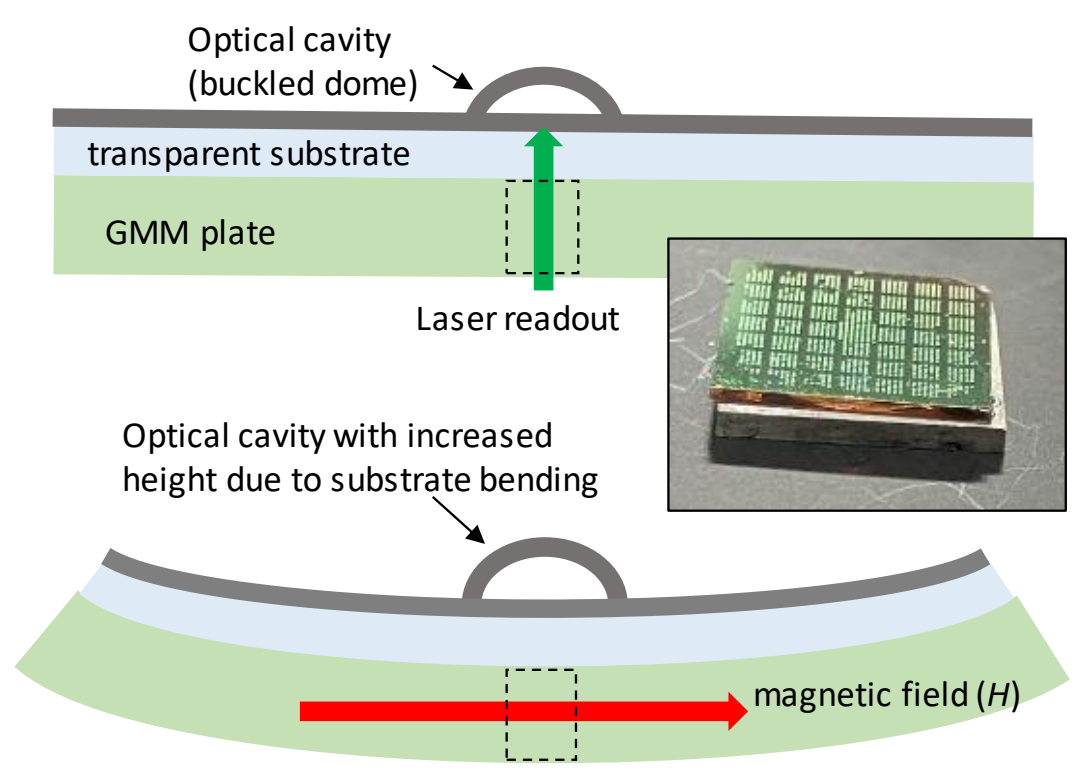

FIG. 1. The principle of the magnetometer is depicted. A chip containing thin-film buckled-dome microcavities is bonded to a GMM. A magnetic field induces strain in the GMM and a bending of the entire assembly. This modulates the stress in the attached thin films and in turn the height of the buckled dome cavity, which can be read out by a laser tuned to the slope of an optical resonance. The inset photograph shows a fused silica chip, ~ 1 $cm^2$ square and 500 μm thick, with arrays of 100 μm diameter buckled dome microcavities visible at the top surface, and bonded to a 1-mm-thick Galfenol plate.

The high-level working principle of our COM is depicted in Fig. 1. A cavity is depicted as lying near the center of its transparent fused silica substrate, which in turn is rigidly bonded to a GMM plate containing a machined hole to allow laser interrogation from below. While many previous COMs [10-12,14,15,17] employed Terfenol-D as the GMM, we used Galfenol [23,24]. Galfenol has lower saturation strain ($\Delta\varepsilon_{max}$ ~ 250 ppm, versus ~ 2000 ppm for Terfenol-D), but a higher piezomagnetic coefficient ($\Delta\varepsilon/\Delta H$ > 1 ppm/Oe)[16,25]. Combined with its superior 'soft magnetic' and mechanical properties discussed further below and in the SI file, this makes Galfenol a better choice for high-frequency sensing applications.

First neglecting the thin films, the assembly in Fig. 1 can be viewed as a bilayer plate system. A magnetic field creates a mismatched strain, which in turn induces a bending response that can be treated using Timoshenko's beam theory [26]. For simplicity, and consistent with the range of values reported for Galfenol [23,25], we assumed an equivalent Young's modulus of 70 GPa for both the fused silica and the GMM plate. For a thickness ratio $t_G/t_{FS} = 2$ (consistent with our fabricated assembly), the induced curvature can then be estimated as [27]:

$$\Delta\kappa \approx \frac{1.3 \cdot \Delta\varepsilon}{T} \quad , \tag{1}$$

where $T = t_{FS} + t_G$ is the total thickness of the bilayer and $\Delta\varepsilon$ is the field-induced strain in the GMM. Using $T = 1.5$ mm and $\Delta\varepsilon/\Delta H \sim 1$ ppm/Oe $\sim 1.25 \times 10^{-6}$ m/A, we can then approximate $\Delta\kappa/\Delta H \sim 1.1 \times 10^{-5}$ $A^{-1}$ or $\Delta\kappa/\Delta B \sim 8.8$ $m^{-1}$ $T^{-1}$ for $B$ in air.

As mentioned, this plate bending induces a stress $\Delta\sigma$ in the attached thin films. As detailed in the supplementary information file, we used the well-known Stoney's equation [28] to relate substrate curvature to film stress, and combined this with our previous treatment [29] of the relationship between changes in film stress and buckled dome cavity height, arriving at the following approximation:

$$\frac{\Delta\delta}{\Delta\kappa} \approx 0.13 \cdot \frac{a^2}{\delta} \cdot \frac{t_{FS}^2}{t_f} \cdot \frac{E_{FS}}{E_f} \cdot \frac{\left(1-\nu_f^2\right)}{(1-\nu_{FS})} \quad . \tag{2}$$

Here, $a$ and $\delta$ are the radius and peak height of the buckled dome cavity, and $t$, $E$, and $\nu$ are the thickness, Young's modulus, and Poisson's coefficient, respectively, of the fused silica substrate or the thin-film stack (i.e., treated as an effective medium) as indicated by the subscripts. This change in height of the cavity produces a proportionate shift in the resonant wavelength, $\Delta\lambda = \lambda \cdot (\Delta\delta/\delta)$, which can be sensitively read out by a laser. For the domes studied below, $a \sim 50$ μm, $\delta \sim 2.4$ μm, and $t_f \sim 1.5$ μm. Furthermore, using $E_f \sim E_{FS} \sim 70$ GPa and $\nu_f \sim \nu_{FS} \sim 0.2$ [29,30], we arrive at $\Delta\delta/\Delta\kappa \sim 3 \times 10^{-5}$ $m^2$. Combining with the results above, it follows that $\Delta\delta/\Delta B \sim (\Delta\delta/\Delta\kappa) \cdot (\Delta\kappa/\Delta B) \sim 2.6 \times 10^{-4}$ m/T.

The sensitivity, or noise-equivalent-magnetic field (NEM), can be expressed [10]:

$$NEM = \frac{S_X}{\left(\Delta\delta/\Delta B\right)} \quad , \tag{3}$$

where $S_x$ is the displacement sensitivity of the optomechanical sensor. We've previously shown [22] that our 100-μm-diameter buckled dome cavities have a displacement sensitivity $S_{X,TH} \sim 4 \times 10^{-17}$ $m/Hz^{1/2}$ in the thermomechanical limit. Using Eq. 3 and the results from above, this implies a potential for $NEM_{TH} \sim 0.15$ $pT/Hz^{1/2}$ if sensing at the thermal limit can be achieved. As shown below, however, other noise sources, especially laser frequency noise, set the effective displacement noise floor ~ 1 - 3 orders of magnitude higher than $S_{X,TH}$ in the frequency range of interest here, so that a sensitivity projection in the pT to hundreds of pT range is more realistic. It should also be noted that the model neglects details such as imperfect bonding between the chip and GMM plate. Moreover, it is essentially a DC model and does not take into account the vibrational resonances of the assembly. Nevertheless, it provides insight and its predictions are in order-of-magnitude agreement with the experimental results below. Additional discussion is provided in the SI file.

Prototype devices (see Fig. 1) were assembled by using a cyanoacrylate glue to bond fused silica chips (~ 1 cm × 1 cm × 500 μm), containing arrays of 100-μm-diameter buckled dome cavities, to custom machined Galfenol plates (~ 1 cm × 1 cm × 1 mm, with a central hole ~ 2 mm in diameter, Lanzhou Jianxing Commercial and Trading Co., China). Glued chip/GMM assemblies were tested in a custom-built Helmholtz coil setup with slots to house permanent rare-earth magnets for bias purposes, and a pair of 60-turn, ~ 1.6 cm diameter coils wound from 28 gauge magnetic wire. The coils were connected in series with a 1 kΩ resistor and the circuit was driven by a signal generator (Rigol DG1022Z), while an oscilloscope connected across the resistor was used to measure the current. The magnetic field produced by the coils was calibrated at low frequency using a commercial Hall-probe magnetometer (Dexinmag #DX-160), and this *I-B* calibration was assumed to hold at the higher frequencies studied below. This ignores the potential for Eddy current losses in the coils at high frequencies, and thus is expected to slightly over-estimate the applied fields. Additional discussion is provided in the SI file.

For all AC measurements below, the magnetometer was biased by a DC magnetic field from a pair of rare-earth magnets spaced by ~ 2.5 cm and collinear with the coil setup. These magnets produced a field of ~ 40 mT at the chip position, verified using the Hall-probe. By trial and error using different numbers of magnets, this field level was found to yield a nearly optimized AC response. This is consistent with expectations for polycrystalline Galfenol, which exhibits a saturation strain at ~ 500-1000 Oe (~50-100 mT in air) [23-25].

With a chip/GMM assembly placed in the coil setup, a fiber-optic-coupled lens (OZ optics) was used to focus light from a tunable 1550-nm-range laser (Santec TSL 710) onto a selected buckled-dome cavity. Reflected light was passed to a high-speed (10 Msamples/sec), digitizing photodetector (Ultracoustics) via a circulator. We first assessed the noise contributions to the cavity readout in the absence of an applied magnetic field, as shown in Fig. 2a. Here, curves are plotted as effective displacement noise, $S_X(f) = \sqrt{S_{XX}(f)}$, using $S_{XX}(f) = S_{PP}(f)/\alpha$, where $\alpha$ is a conversion factor between displacement and optical power spectral density [22]. For the cavity and interrogation power used here, $\alpha \sim 10^{11}\ W^2/m^2$ (see the SI file for details). The blue solid curve is the measured noise spectrum, which has two dominant peaks at ~ 2.4 and ~ 4 MHz associated with the two lowest-order radially symmetric vibrational modes of the buckled microcavity [22]. We used a dual harmonic oscillator model to fit to these features, yielding an estimate for the thermomechanical noise spectrum (red solid curve). Shot noise (estimated from the average received power of ~ 100 μW) and dark noise (measured) are also plotted, and lie below the TM noise over most of the range. The sum of TM, shot, and dark noise contributions is also shown (dotted purple curve), revealing that below ~ 500 kHz there is an additional and dominant noise source. We have verified that this is overwhelmingly due to laser phase noise [12,15], converted to intensity noise in the process of interrogating the high-$Q$ Fabry-Perot cavity.

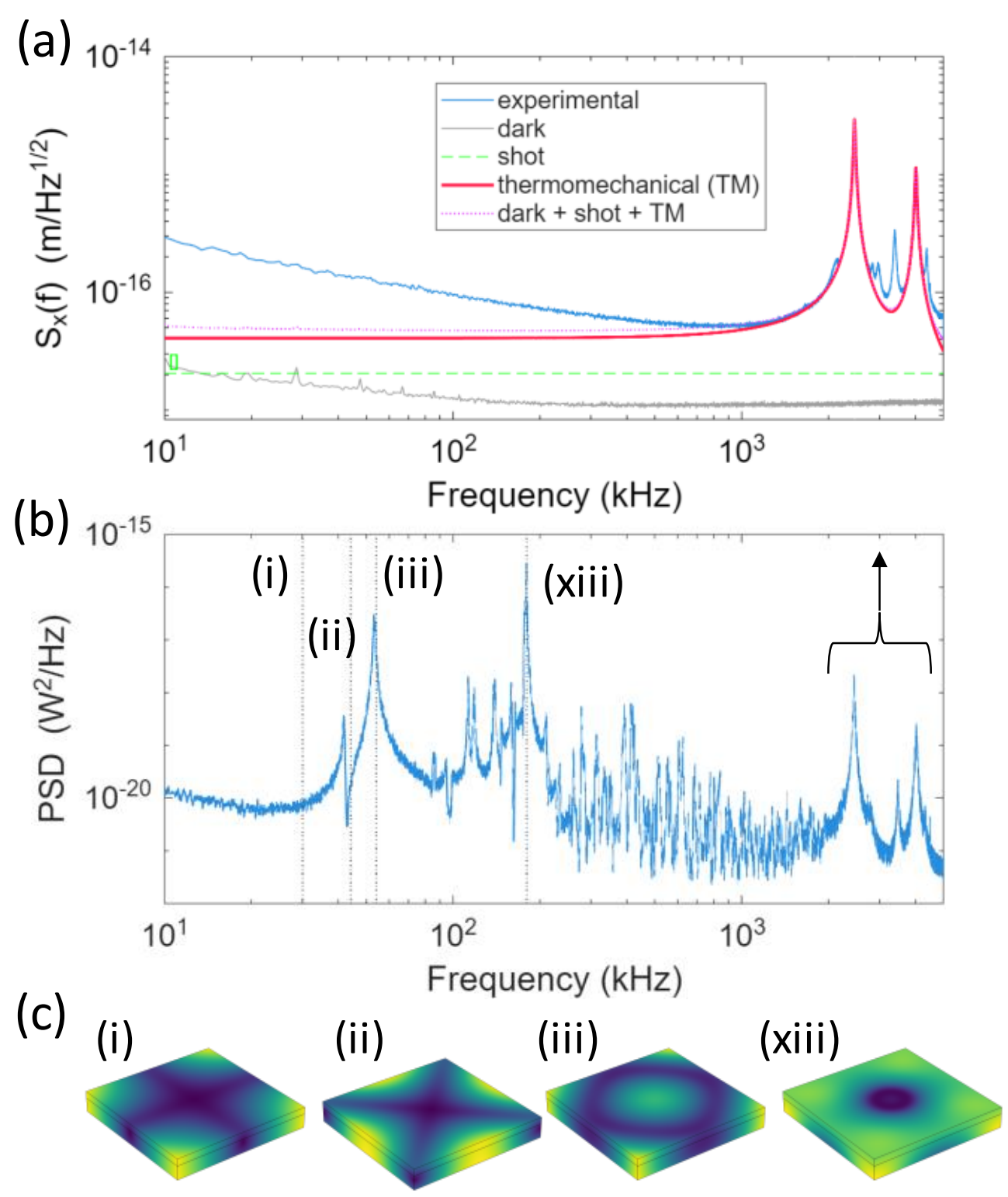

FIG. 2. (a) Noise contributions plotted as effective displacement spectral density. The gap between the measured noise (blue solid curve) and the sum of TM, shot, and dark noise (purple dotted curve) is attributed to laser phase noise. (b) Raw PSD measured with the Helmholtz coil setup driven by the white-noise function of the signal generator. The vertical dashed lines indicate the three lowest and also the 13th numerically predicted eigenfrequencies of the chip/GMM assembly with free boundary conditions. (c) Spatial mode profiles for the eigenmodes corresponding to the labeled frequencies in part b, as predicted using a numerical solver (COMSOL).

We next drove the coils using the pseudo-white noise function of the signal generator, and Fig. 2b shows a typical resulting spectrum. In addition to the MHz-range vibrational modes of the buckled dome, the spectrum contains an array of resonant features lying in the ~ 40 kHz to ~ 2 MHz range. We attribute these to the vibrational modes of the chip/GMM assembly, which is mounted 'freely' in the setup, supported only by small posts near its corners. We used a numerical solver (COMSOL Solid Mechanics Module) to simulate the assembly, under the assumption of a rigidly bonded interface. Details can be found in the SI file. The dashed vertical lines in Fig. 2b indicate the three lowest-order resonance frequencies predicted by the model. Also indicated is the 13$^{th}$ eigen-frequency, since it aligns very well with the strongest observed experimental peak at ~ 179 kHz.

The numerically predicted spatial profiles of these eigenmodes are shown in Fig. 2c. Modes (i)-(iii) are essentially those of a square plate with free boundaries [31]. The first two modes have nodes near the center, and their flexural motion is presumably less well coupled to the interrogated cavity. In particular, the lowest-order mode of a free square plate is a torsional 'flapping' mode characterized by opposite corners moving in unison and with very little displacement or bending near the center [22,31], explaining the absence of this feature in the experimental spectrum. Since low-frequency magnetic fields predominately couple to that mode in our prototype magnetometer, this reduces the sensitivity in the range below ~ 50 kHz as discussed further below.

Mode (iii), on the other hand, has a centro-symmetric anti-node, and a strong resonant response near its predicted eigen-frequency (~ 54 kHz) was consistently observed across multiple chip/GMM assemblies (see the SI file for other examples) and regardless of the particular buckled dome cavity interrogated on a given assembly. Typically, the strongest response was observed at a higher frequency, for example at ~ 179 kHz for the case shown in Fig. 2. This frequency aligns very well with the eigenmode labeled as (xiii), which has a surface displacement node near the center but likely also induces strong bending of the chip in the vicinity of the interrogated domes. Other assemblies exhibited their strongest response for nearby eignenmodes (see the SI file for examples), likely due to variations in the geometry of the assemblies and the location of a particular dome cavity chosen for interrogation.

A typical response to a harmonic applied field, in this case ~ 0.06 μT at the most strongly coupled mechanical resonance frequency of 179 kHz, is shown in Fig. 3a. A strong response at the excitation frequency is observed, with SNR > $10^7$. Relatively weak harmonic overtones, more than 5 orders of magnitude below the fundamental, are visible at 358 and 547 kHz, and are attributable to readout nonlinearities for the tuned-to-slope technique [19]. The peak PSD at the fundamental frequency is shown in the inset plot, demonstrating a linear response up to ~ 2.5 μT. It is worth noting that more sophisticated readout schemes that track the optical resonance frequency could be used to greatly increase the linear dynamic range.

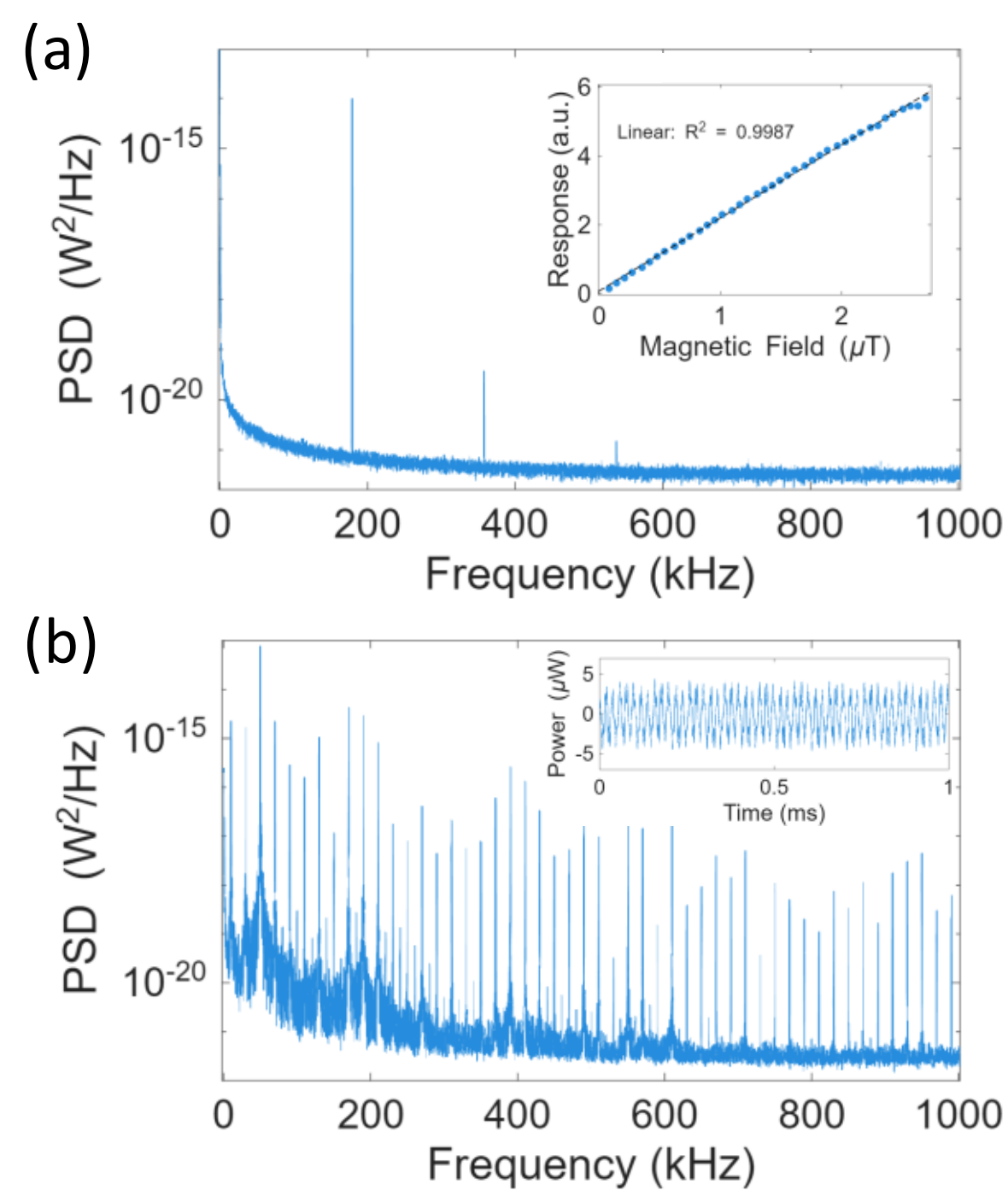

FIG. 3. (a) The spectral response with the COM driven by a 179 kHz, ~ 0.06 μT sinusoidal *B*-field. Inset: plot of peak response at frequency of 179 kHz versus applied magnetic field. (b) The spectral response with the COM driven by a 10 kHz square wave, with similar peak field as used in part a. Inset: the corresponding time-domain signal.

The spectral response to a 10 kHz square-wave excitation, with similar peak driving voltage to that used in part a., is shown in Fig. 3b. Along with the fundamental, the spectrum comprises the series of odd-order harmonic frequencies expected for a square wave, extending into the MHz range. The corresponding time-domain signal, with peak-to-peak optical power excursions lying within the linear readout range of the cavity, is shown in the inset plot. The signal is not a faithful representation of the input square wave, due to linear distortion [16] arising from the frequency-dependent response (see Fig. 2b and Fig. 4a below). Nevertheless, the rich harmonic spectrum and sharp features in the time-domain plot both demonstrate the broadband character of the magnetometer. In principle, the distorted signal could be corrected by deconvolution of the magnetometer spectral response.

Finally, we evaluated the magnetometer response by stepping the frequency of a harmonic excitation at fixed drive voltage. The current delivered to the coil was measured at each step and used to estimate the applied magnetic field. A raw response spectrum in the 1 kHz to 1 MHz range is shown in Fig. 4a. Note that this response is essentially a higher-SNR version of the plot obtained using white-noise excitation (Fig. 2b), revealing the same resonant features associated with the chip/GMM assembly. Following the approach described elsewhere, we estimated the magnetometer sensitivity as $NEM(f) = B(f)/\{SNR(f)\cdot RBW\}^{1/2}$, where $B$ is the applied magnetic field estimated from the measured current and $RBW$ (1 Hz here) is the resolution bandwidth of the measurement. A typical result is plotted in Fig. 4b, demonstrating a broadband sensitivity better than ~ 1 nT/Hz$^{1/2}$ over most of the range, and a peak sensitivity ~ 2 pT/Hz$^{1/2}$ at ~ 179 kHz which is similar to the best previously reported values for a COM [14,15]. In keeping with the discussion above, the sensitivity below ~ 40 kHz is reduced due to relatively poor coupling of the fundamental vibrational mode of the assembly to the near-center interrogated dome. This could be addressed by changing the geometry or the mounting and boundary conditions of the assembly, but is left for future work.

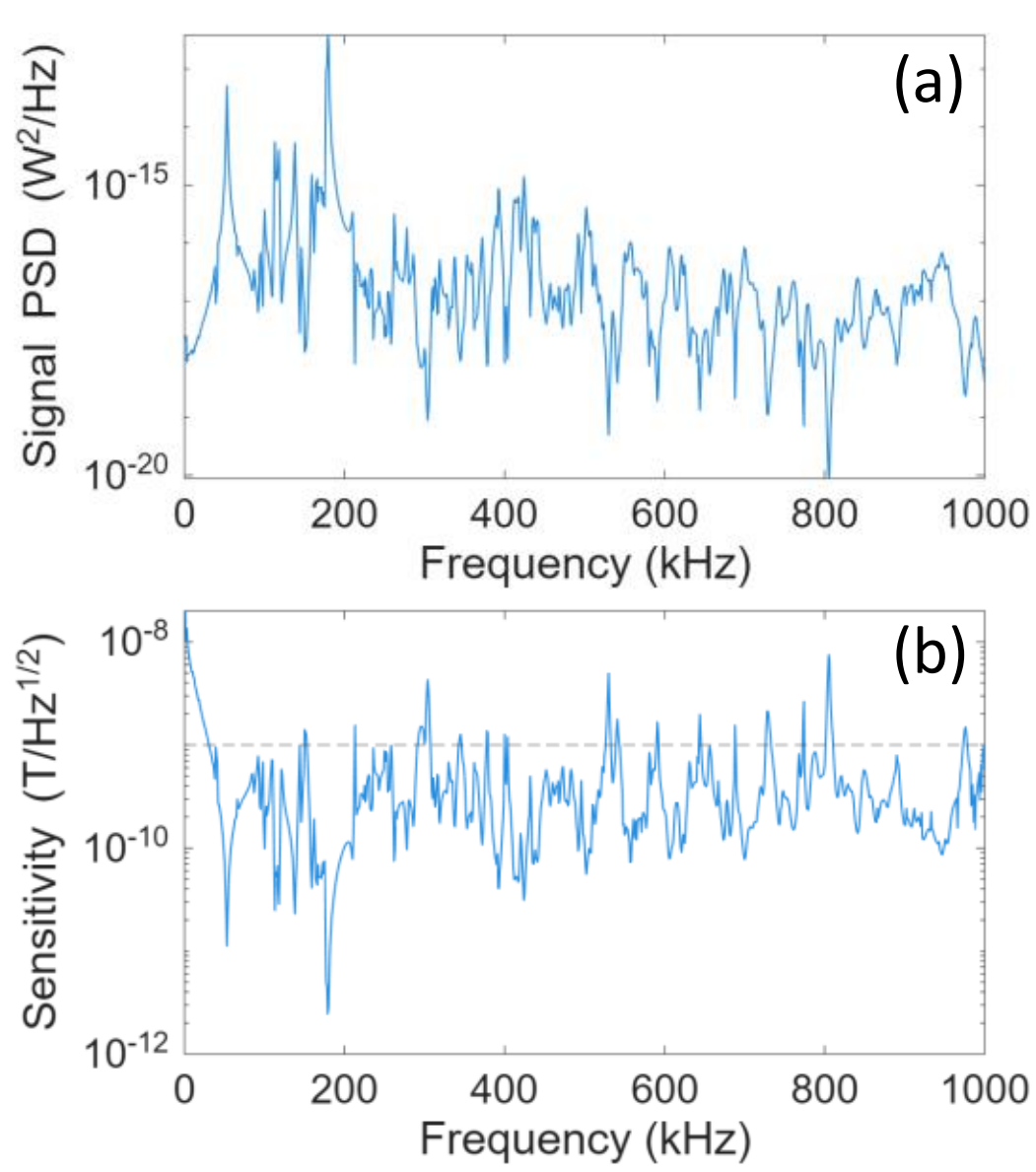

FIG. 4. (a) Raw signal response versus frequency with constant supply voltage. (b) Noise-equivalent magnetic field (NEM) versus frequency extracted from the sweep in part a. The dashed horizontal line indicates 1 nT/Hz$^{1/2}$.

In summary, we described an optomechanical magnetometer capable of detecting magnetic fields with single-digit-picotesla sensitivity in the kHz range, in a room-temperature, unshielded laboratory environment, and using a straightforward laser interrogation setup. Broadband (over the ~ 1 kHz to 1 MHz range) sensitivity better than 10 $nT/Hz^{1/2}$ was demonstrated. The sensitivity is currently limited in large part by laser phase noise. In principle, this noise could be reduced by use of a narrower linewidth laser [20] or by using an interferometric phase stabilization approach [12]. Other strategies include frequency mixing of the target signal up to a frequency range with lower phase noise [14], or using a digital bandpass filter [13] or lock-in technique [32] to suppress noise in the case of a known, narrowband signal. We project that, with the suppression of laser phase noise, the current assembly could enable thermomechanical-noise-limited sensitivity < 1 $pT/Hz^{1/2}$ over a bandwidth extending from near DC to several MHz. Moreover, there is potential for additional optimization of the assembly by improving the bonding method and by adjusting the geometry of the various components. We hope to explore this, and the reduction of laser phase noise, in future work.

## SUPPLEMENTARY MATERIAL

See supplementary material for extra information concerning the experimental setup, the buckled dome cavities, signal processing details, COMSOL simulations of chip/GMM vibrational modes, and additional results.

## ACKNOWLEDGEMENTS

The authors would like to acknowledge the following funding sources which supported this work:

## REFERENCES

1. J. Lenz and A. S. Edelstein, "Magnetic sensors and their applications," IEEE Sensors J. **6**(3), 631-649 (2006).
2. K.-M. C. Fu, G. Z. Iwata, A. Wickenbrock, and D. Budker, "Sensitive magnetometry in challenging environments," AVS Quantum Sci. **2**, 044702 (2020).
3. D. Murzin, D. J. Mapps, K. Levada, V. Belyaev, A. Omelyanchik, L. Panina, and V. Rodionova, "Ultrasensitive magnetic field sensors for biomedical applications," Sensors **20**, 1569 (2020).

[4] J. S. Bennett, B. E. Vyhnalek, H. Greenall, E. M. Bridge, F. Gotardo, S. Forstner, G. I. Harris, F. A. Miranda, and W. P. Bowen, "Precision magnetometers for aerospace applications," Sensors **21**, 5568 (2021).
[5] D. Budker and M. Romalis, "Optical magnetometry," Nature Phys. **3**, 227-234 (2007).
[6] see https://twinleaf.com/ and https://quspin.com/, accessed July 24, 2026.
[7] see https://sbquantum.com/, accessed July 24, 2026.
[8] H. Kumar, S. Dasika, M. Mangat, S. Tallur, and K. Saha, "High dynamic-range and portable magnetometer using ensemble nitrogen-vacancy centers in diamond," Rev. Sci. Instrum. **95**, 075002 (2024).
[9] J. F. Barry, M. H. Steinecker, S. T. Alsid, J. Majumder, L. M. Pham, M. F. O'Keeffe, and D. A. Braje, "Sensitive ac and dc magnetometry with nitrogen-vacancy-center ensembles in diamond," Phys. Rev. Appl. **22**, 044069 (2024).
[10] S. Forstner, S. Prams, J. Knittel, E. D. van Ooijen, J. D. Swaim, G. I. Harris, A. Szorkovszky, W. P. Bowen, and H. Rubinsztein-Dunlop, "Cavity optomechanical magnetomer," Phys. Rev. Lett. **108**, 120801 (2012).
[11] S. Forstner, E. Sheridan, J. Knittel, C. L. Humphreys, G. A. Brawley, H. Rubinstein-Dunlop, and W. P. Bowen, "Ultrasensitive optomechanical magnetometer," Adv. Mater. **26**, 6348-6353 (2014).
[12] C. Yu, J. Janousek, E. Sheridan, D. L. McAuslan, H. Rubenstein-Dunlop, P. K. Lam, Y. Zhang, and W. P. Bowen, "Optomechanical magnetometry with a macroscopic resonator," Phys. Rev. Appl. **5**, 044007 (2016).
[13] J. Li, E. Xing, T. Jia, T. Xing, G. Xing, D. Wang, J. Rong, L. Li, and J. Liu, "Ultra-sensitive magnetic sensor based on resonator with asymmetric wedge structure," IEEE Sensors J. **24**(8), 12244-12250 (2024).
[14] A.-N. Xu, Y. Li, X. Li, B. Liu, and Y.-C. Liu, "Subpicotesla optomechanical magnetometry," Phys. Rev. Lett. **133**, 153601 (2024).
[15] S. Yang, Q. Zhang, L. Yang, H. Liu, Q. Wang, P. Zhang, H. Shen, and Y. Li, "Optomechanical sensor network with fiber Bragg gratings," Nat. Commun. **16**, 10795 (2025).
[16] Z.-G. Hu, Y.-M. Gao, J.-F. Liu, H. Yang, M. Wang, Y. Lei, X. Zhou, J. Li, X. Cao, J. Liang, C.-Q. Hu, Z. Li, Y.-C. Lau(, J.-W. Cai, and B.-B. Li, "Picotesla-sensitivity microcavity optomechanical magnetometery," Light Sci. Appl. **13**, 279 (2024).
[17] Q. Zhang, S. Du, S. Yang, Q. Wang, and Y. Li, "Membrane optomechanical magnetometer," Opt. Express **33**(4), 8774-8781 (2025).
[18] B.-B. Li, L. Ou, Y. Lei, and Y.-C. Liu, "Cavity optomechanical sensing," Nanophotonics **10**(11), 2799-2832 (2021).
[19] G.J. Hornig, K.G. Scheuer, E.B. Dew, R. Zemp, and R.G. DeCorby, "Ultrasound sensing at thermomechanical limits with optomechanical buckled-dome microcavities," Opt. Express **30**(18), 33083 (2022).
[20] F. Zhou, Y. Bao, R. Madugani, et al., "Broadband thermomechanically limited sensing with an optomechanical accelerometer," Optica 8(3), 350–356 (2021).
[21] K.G. Scheuer, and R.G. DeCorby, "Air-coupled ultrasound using broadband shock waves from piezoelectric spark igniters," Appl. Phys. Lett. **125**, 082202 (2024).
[22] K. G. Scheuer, F. B. Romero, and R. G. DeCorby, "Spectroscopy of substrate thermal vibrational modes using an optomechanical sensor," Opt. Mater. Express **14**(7), 1793-1802 (2024).
[23] A. Orr, J. Downing, H. Kim, J. Paglione, J. Cozzo, S.-M. Na, and A. Flatau, "Influence of shape anisotropy and temperature on magnetostrictive behavior in single-crystal Galfenol alloys," IEEE Trans. Magn. **53**(11), 2503804 (2017).
[24] Y. Chen, T. Yang, J. Gou, R. Qiao, Y. Zhang, and T. Ma, "Synergistic effects of magnetic domain re-alignment and precipitation on achieving large magnetostriction in random polycrystalline Fe-Ga alloys," Materials & Design **256**, 114324 (2025).
[25] J. R. Downing, S.-M. Na, and A. Flatau, "Compressive pre-stress effects on magnetostrictive behaviors in highly textured Galfenol and Alfenol thin sheets," AIP Advances **7**, 056420 (2017).

[26] S. Timoshenko, "Analysis of bi-metal thermostats," J. Opt. Soc. Am. **11**(3), 233-255 (1925).
[27] A. I. Egunov, J. G. Korvink, and V. A. Luchnikov, "Polymethylsiloxane bilayer films with an embedded spontaneous curvature," Soft Matter **12**, 45 (2016).
[28] S. J. McKeown, X. Wang, X. Yu, and L. L. Goddard, "Realization of palladium-based optomechanical cantilever hydrogen sensor," Microsyst. & Nanoeng. **3**, 16087 (2017).
[29] M. H. Bitarafan, H. Ramp, T. W. Allen, C. Potts, X. Rojas, A. J. R. MacDonald, J. P. Davis, and R. G. DeCorby, "Thermomechanical characterization of on-chip buckled dome Fabry-Perot microcavities," J. Opt. Soc. Am. B **32**(6), 1214-1220 (2015).
[30] K. G. Scheuer, F. G. Romero, and R. G. DeCorby, "Coupling the thermal acoustic modes of a bubble to an optomechanical sensor," Microsyst. Nanoeng. **10**, 204 (2024).
[31] T. D. Rossing and D. A. Russell, "Laboratory observation of elastic waves in solids," Am. J. Phys. **58**(12), 1153-1162 (1990).
[32] J. Pretula, N. Shaw, A. Chen, K. G. Scheuer, and R. G. DeCorby, "Resonant ultrasound spectroscopy detection using a non-contact ultrasound microphone," Sensors **25**, 6154 (2025).